\RequirePackage[switch,columnwise]{lineno}
\documentclass[twocolumn,twoside,slac_two]{revtex4}
\usepackage{graphicx}
\usepackage{fancyhdr}
\usepackage{hyperref}
\hypersetup{
    colorlinks=true,
    urlcolor=magenta,
}
\def\matA{{\ensuremath{\underline{\underline{A}}}}}
\def\vecg{{\ensuremath{\vec{g}}}}
\def\vecf{{\ensuremath{\vec{f}}}}
\def\vecb{{\ensuremath{\vec{b}}}}
\def\be{\begin{equation}}
\def\ee{\end{equation}}
\def\bea{\begin{eqnarray}}
\def\eea{\end{eqnarray}}
\def\Depth{{\tt{Depth}}}
\def\Esurf{\ensuremath{\mathrm{\tt{E}}_\mu}}
\def\dEdx{\ensuremath{\tt{dE/dx}_{SplineMPE BINS}}}
\def\Conv{SIBYLL-2.1}
\def\Charm{ERS}
\def\Unfl{Unflavored}

\begin{document}

\title{LaTex Template for ECRS 2016}

%

\title{Development of a Machine Learning Based Analysis Chain for the Measurement of Atmospheric Muon Spectra with IceCube}
\author{Tomasz Fuchs for the IceCube Collaboration\footnote{\protect\url{http://icecube.wisc.edu}}}
\affiliation{Technical University Dortmund, Experimental Physics V, \\
D-44221 Dortmund, Germany}

\begin{abstract}
High-energy muons from air shower events detected in IceCube are selected using state of the art machine learning algorithms.
Attributes to distinguish a HE-muon event from the background of low-energy muon bundles are selected using the mRMR algorithm and the events are classified by a random forest model.
In a subsequent analysis step the obtained sample is used to reconstruct the atmospheric muon energy spectrum, using the unfolding software TRUEE.
The reconstructed spectrum covers an energy range from $10^4$\,GeV to $10^6$\,GeV.
The general analysis scheme is presented, including results using the first year of data taken with IceCube in its complete configuration with $86$ instrumented strings.
\end{abstract}

\maketitle

\thispagestyle{fancy}


\section{Introduction}
IceCube is a cubic kilometer detector array located at the geographic South Pole.
Its $5160$ Digital Optical Modules (DOMs) are used to detect secondary muons produced either in neutrino interactions with ice or bedrock, or in cosmic ray air showers.
\renewcommand*{\thefootnote}{\alph{footnote}}
The combination of the conventional (i.e. produced in pion and kaon decays) atmospheric and astropysical neutrino flux dominates over the expected prompt component which is produced mainly by charmed hadron decays in the atmosphere.
Therefore an accurate measurement of the prompt flux magnitude is a difficult task for a large-volume neutrino detector.
Since the energy spectrum of atmospheric muons has no astrophysical component, this flux can be studied to determine the magnitude of the prompt flux which is produced in leptonic decays of charmed hadrons and
unflavored mesons.

In general high-energy muons from air showers are accompanied by a bundle of low-energy muons.
Therefore, the detection of HE-muons within a muon bundle is a challenging task as the IceCube detector lacks the spatial resolution to resolve individual muons within a muon bundle.
To analyze the atmospheric muons we developed a data mining based analysis scheme that selects algorithmically from a large pool of reconstructed event properties the most powerful ones for distinguishing signal and background.
A similar approach was used to analyze atmospheric neutrinos\cite{tim} in IceCube.

The developed analysis chain was used to reconstruct the spectrum of high-energy muons reaching the IceCube in-ice detector.
For this analysis a HE-muon is defined as a muon which fulfills the following condition \be E_{\mu,\mathrm{HE}} \geq 0.5 E_{\mathrm{bundle},\mathrm{tot.}}. \label{eq:Emax}\ee
$E_{\mu,\mathrm{HE}}$ describes the energy of the HE-muon and $E_{\mathrm{bundle},\mathrm{tot.}}$ the total bundle energy including the \mbox{HE-muon}.
Since the muons within the bundle cannot be resolved spatially, high-energy muons can only be identified indirectly from the relation between catastrophic and continuous energy losses, and separation from background is challenging.

Our analysis chain has the following three parts: a) the selection of event properties (hereafter referred to as attributes) to be used in the analysis, b) the classification of events into signal and background using these attributes, and c) the reconstruction of  the spectrum of the HE-muons  at the surface.

\section{Attribute Selection}
Choosing the right attributes is a crucial step for every analysis.
For this proceeding every value which is measured or reconstructed for every single detected or simulated event is referred to as an attribute.
From the pool of all available attributes we select algorithmically those that agree between data and simulation using a comparison of the integrated distributions.
This is also a criterion in the attribute selection since only well simulated attributes can be used to draw conclusions for measured events.
For following steps it is common to manually select attributes based on experience and expert knowledge.
In this analysis the large dimensionality of the set of attributes renders this procedure impossible.

An algorithmic approach like the mRMR algorithm\cite{ding05} is necessary.
In the mRMR algorithm the relevance $D$ and the redundancy $R$ of a specific set of attributes is analyzed.
In a specific attribute set the relevance $D$ describes the correlation of each attribute to the signal and background events whereas the redundancy $R$ quantifies the correlation between the attributes in this set.
In the mRMR algorithm a maximization of the equation \be\Phi(D,R)= D-R\label{eq:Phi}\ee is performed.
The attribute set is built up iteratively by adding a single attribute which maximizes $\Phi(D,R)$ in equation \ref{eq:Phi} in each step.
The advantage of this procedure is the possibility to determine the best attributes which are able to classify the data for any size of the attribute set.
In this analysis the final attribute set uses $30$ attributes which proved to be a good tradeoff between execution time and separation quality.

\section{Event Classification}
\begin{figure}[h]
\centering
\begin{minipage}{0.45\textwidth}
\includegraphics[trim = 0px 0px 0px 0px, clip, width=\textwidth]{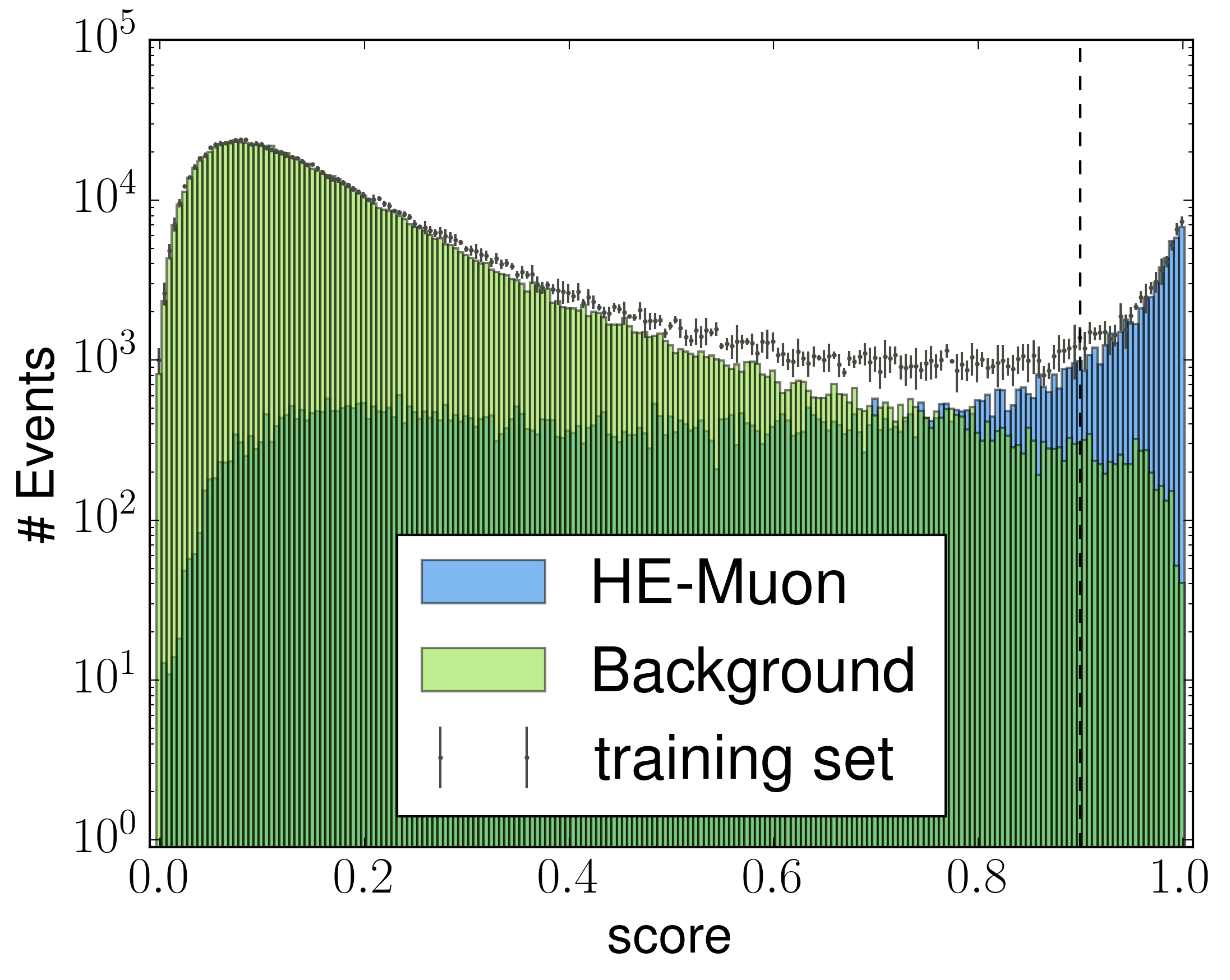}
\caption{Random Forest score for the training and the test set for the separation of HE-muons and events without high-energy muons.
Additionally the chosen score limit of $0.9$ is shown.
} \label{ConfMC}
\end{minipage}\hfill
\begin{minipage}{0.45\textwidth}
\includegraphics[trim = 0px 0px 0px 0px, clip, width=\textwidth]{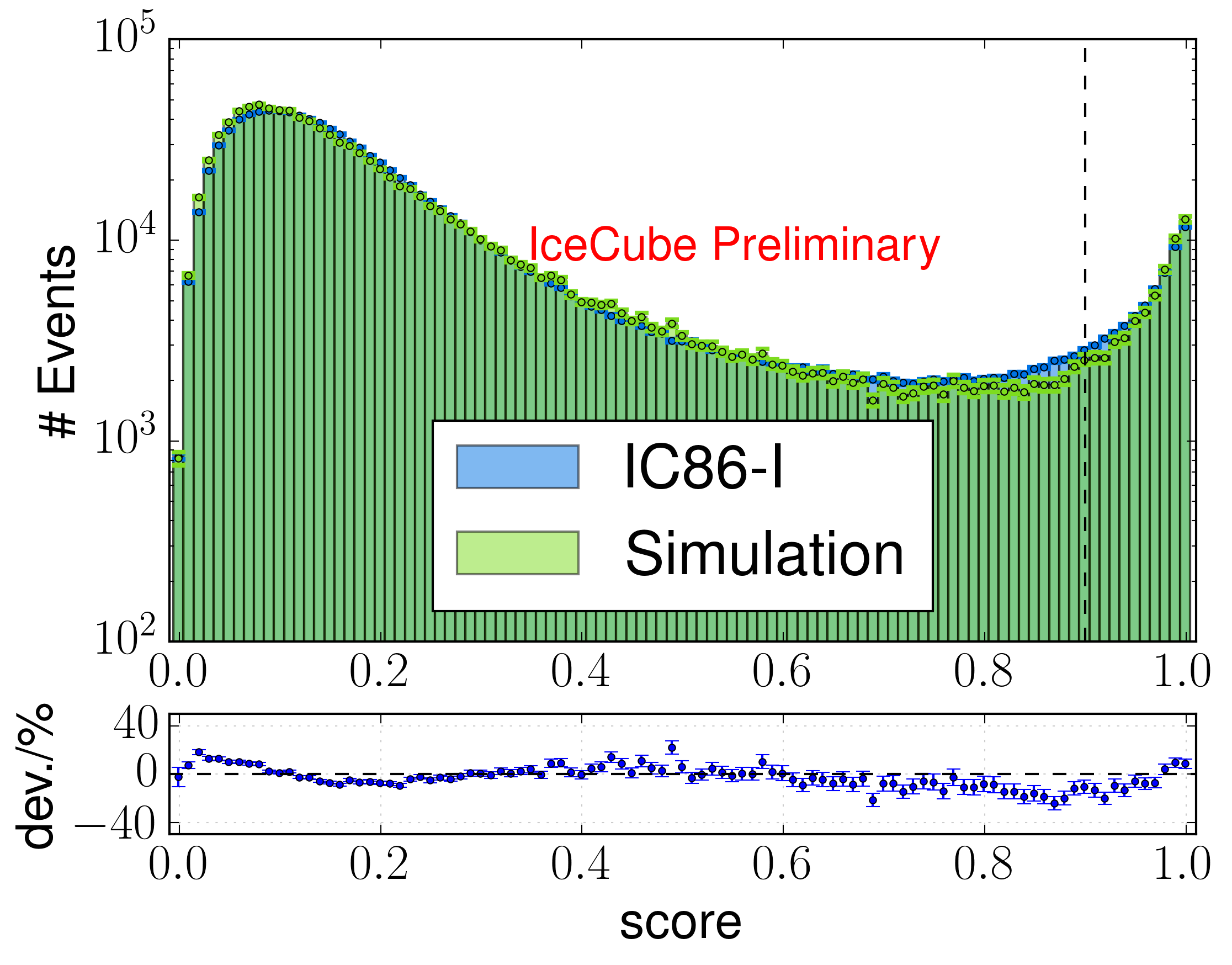}
\caption{
Comparison of the random forest score between simulation and data observed with IceCube ($10$\,\% of available statistics for IC86-I).
Also the chosen score limit of $0.9$ and the deviation between the simulated events and the measured data is shown.} \label{ConfData}
\end{minipage}
\end{figure}

With the selected set of attributes a classification of events into signal and background has to be done.
This can be achieved by splitting the event sample at a certain value of a certain attribute (a so called "straight cut"), but algorithms from computer science like deep neural networks\cite{dl}, boosted decision trees\cite{bdt} or random forests\cite{rfbrei} are increasingly used and show comparable or superior performance.
Here the random forest algorithm is used to perform the selection of HE-muons.
A random forest is an ensemble of single decision trees which select random attributes at every knot and is proven to be more robust against overtraining.
In the case of a two-class classification problem the random forest determines a score which indicates the most likely class affiliation.
Here the score is used to distinguish between signal (HE-muons) and background events (low-energy muon bundles).
Because the random forest builds multiple random decision trees, its score can simply be calculated by the fraction of single trees which classified an event to be signal $T_+$ over the total number of built trees $T$.
\be S = T_+ / T\label{eq:S}\ee
To avoid a misleading classification result by an overtrained model the data to build the model is split into a training and a test set with randomly chosen events.
The results of the model trained on the training set applied to the test set can be seen in Figure \ref{ConfMC}.
Both classes can be distinguished in the score distribution, so it can be used to separate the HE-muons from background events using a straight cut in the score value.

To evaluate the optimal cut values, different performance indicators are used.
The most common indicators are the efficiency and the purity of a potential set of events.
The efficiency describes the fraction of signal events above a specific score value.
The purity is the ratio of signal events to the total number of events above a specific score value.
The requirements on the purity and efficiency are problem dependent so that a general approach is not available.
In this analysis a score cut of $S \geq 0.9$ is used which results in an efficiency of $(40.8\pm 0.6)\,\%$ and a purity of $(93.1\pm 0.4)\,\%$.
The score cut was chosen because for higher cut values the efficiency of the resulting sample drops significantly.

To rule out a possible mismatch between data and simulation the score distribution on simulated events is compared to data events.
In Figure \ref{ConfData} an agreement within $20$\,\% between simulated and analyzed data events is seen. 
Since the data used to train the model is limited there is an uncertainty for the score distributions as well as the efficiency and purity.
The uncertainty for the score, efficiency and purity are estimated using a $5$-fold cross validation.
In a cross validation the training set is split into $n$ parts.
The model is trained on $n - 1$ parts and applied on the left-out part.
This procedure is repeated until every part has been used as a test set.
The variance of performance values of the multiple classifications is used to estimate the uncertainty.

\begin{figure}[h]
\centering
\begin{minipage}{0.5\textwidth}
\includegraphics[trim = 0px 0px 0px 0px, clip, width=\textwidth]{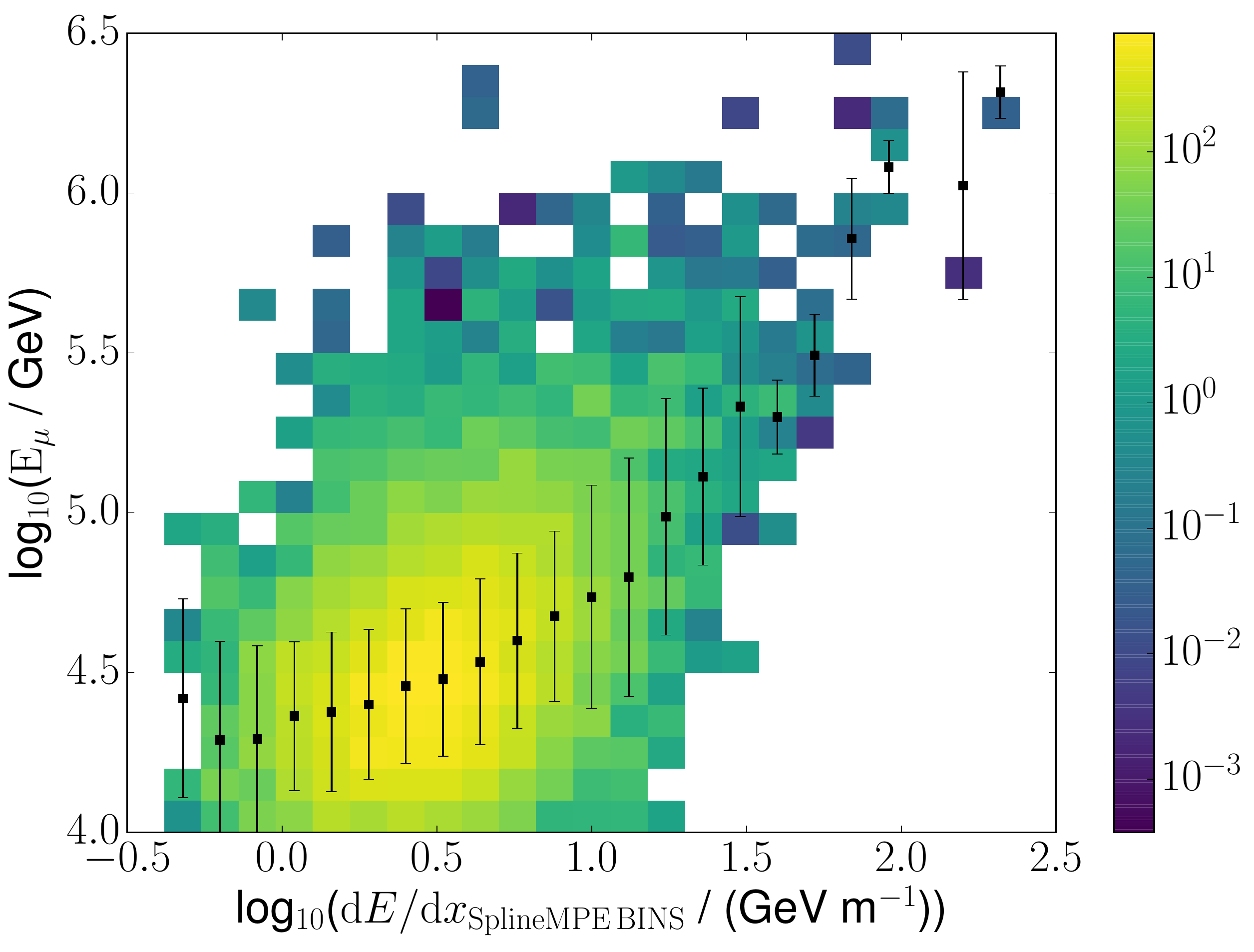}
\caption{Correlation between the surface muon energy and the energy estimator. For each energy estimator bin the mean and the standard deviation of the surface energy are shown.} \label{CorrDeDx}
\end{minipage}\hfill
\begin{minipage}{0.5\textwidth}
\includegraphics[trim = 0px 0px 0px 0px, clip, width=\textwidth]{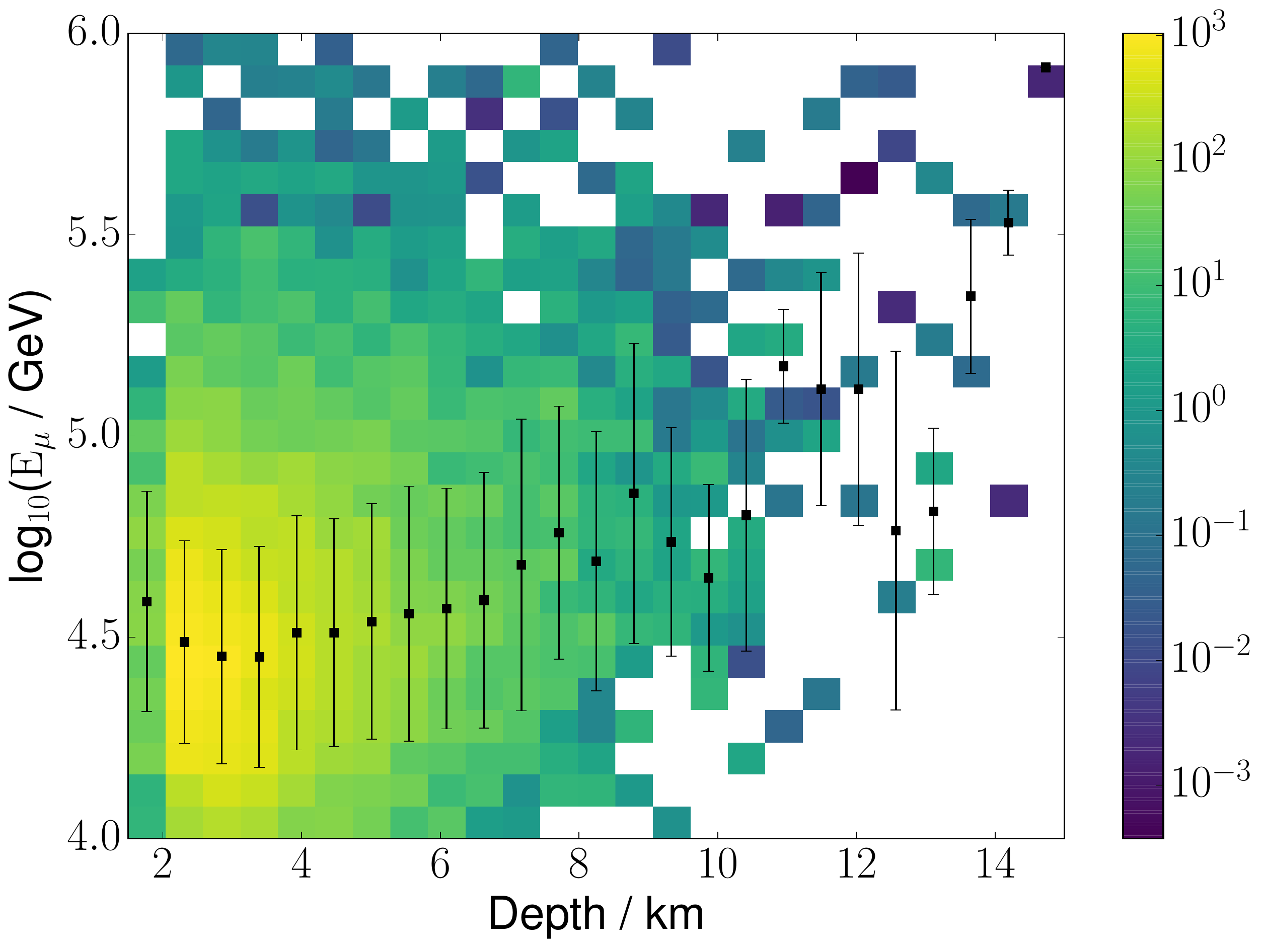}
\caption{Correlation between the surface muon energy and the propagated distance in the ice. For each distance bin the mean and the standard deviation of the surface energy are shown.} \label{CorrDepth}
\end{minipage}
\end{figure}
\section{Unfolding}
The muon energy at the surface cannot be measured directly.
Its distribution has to be reconstructed from observables that can be measured at the detector.
A common method is the forward folding which fits the expected event properties to the data varying parameters of a model of the muon energy spectrum at the surface.
This method has the disadvantage that the resulting spectral shape is limited to the chosen parametrization.
An alternative approach to this is the unfolding method which estimates the surface muon spectrum directly by inverting the response functions which describe the expected event properties for each value of the surface energy of the muon.
This method is based on a Fredholm integral of the second kind which can be discretized to \be \vecg = \matA \vecf + \vecb .\ee
The sought after distribution \vecf\ is folded with a response matrix \matA .
The response matrix \matA\ describes the expected distribution of a specific attribute \vecg\ for each bin in surface muon energy.
A background contribution \vecb\ has to be considered in addition to obtain the expected distribution of \vecf .
The matrix \matA\ is determined from simulated events and the reconstruction of \vecf\ is done using a regularized\cite{tik} likelihood fit with the software TRUEE\cite{truee}.
In the case of HE-muons two attributes are used for the reconstruction of the surface energy.
This surface energy of the muon is determined by the reconstructed energy losses of the muon in the detector and the estimated energy losses occuring during the propagation from the surface to the detector.
To account for their energy loss prior to reaching IceCube, the distance from the surface to the detector boundary is used since no measurement can be performed outside the detector.
The correlation between the surface energy \Esurf\ and the reconstructed energy loss \dEdx\ and the distance from the surface \Depth\ of the events are shown in the Figures \ref{CorrDeDx} and \ref{CorrDepth}. 

\begin{figure*}[h]
\centering
\includegraphics[width=160mm]{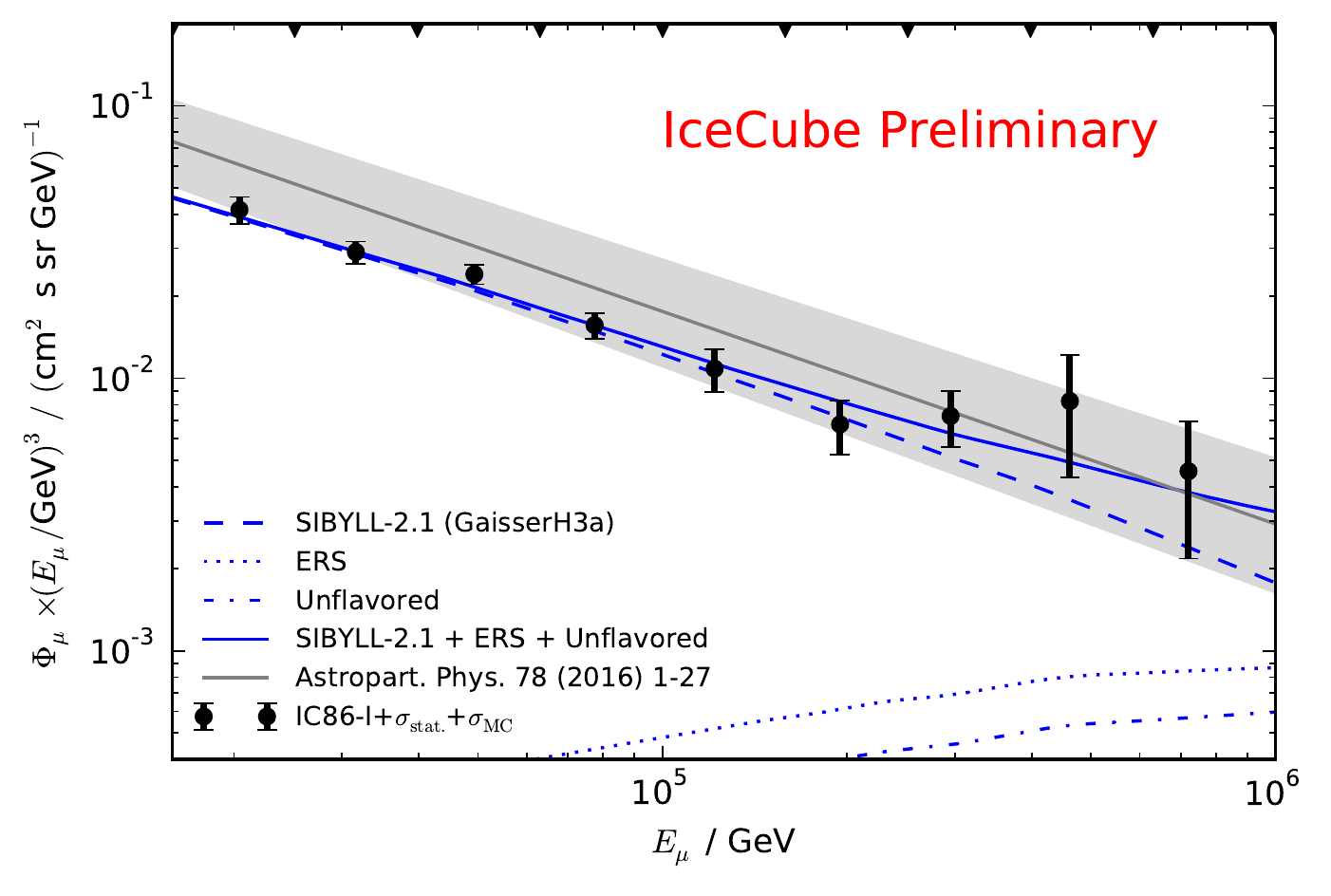}
\caption{Unfolded muon spectrum multiplied with E$^3$ using the IC86-I data. The reconstructed muon spectrum from AP.78 is shown in gray. In blue the theoretic prediction for a conventional part (\Conv \cite{sybillflux}) with the assumed GaisserH3a\cite{h3a} cosmic ray flux model in brackets are shown and a the contribution by a prompt component from charm (\Charm \cite{charm}) and unflavored (\Unfl \cite{unfl}) particles.}
\label{Specfinal}
\end{figure*}

Using these attributes a surface energy spectrum of the HE-muons can be determined and is shown in Figure \ref{Specfinal}.
To enhance the visibility of spectral features of the reconstructed flux the result is multiplied by $(\Esurf/\mathrm{GeV})^3$.
The blue lines in Figure \ref{Specfinal} are theoretic predictions for the flux and are split into a conventional part (using SIBYLL-2.1\cite{sybillflux} as the hadronic interaction model and GaisserH3a\cite{h3a} as the cosmic-ray spectrum) and a prompt contribution from charm (\Charm \cite{charm}) and unflavored (\Unfl \cite{unfl}) particles.
A previous analysis without a machine learning approach published in \cite{patrick} is shown with its uncertainties as a gray band.
The results of this analysis are plotted as black circles.

The uncertainty of this analysis has two contributions. 
A smaller one from the limited statistics of observed events and a larger one from the limited statistics of simulated signal events.
To estimate the uncertainty from the limited simulation events the spectrum was reconstructed using multiple resampled subsets of the total set.
The variance of the spectra was then considered in the estimation of the uncertainty.
This yields a large uncertainty, especially for the higher energies ($E_\mu > 10^5$ GeV).
This is due to the fact that events with more energy are far less often simulated due to the assumed $E^{-2}$ cosmic ray spectrum in the simulation.
\newpage
\onecolumngrid

\twocolumngrid

\section{Conclusion and Future Perspectives}
The described machine learning based analysis chain proved that it can be used to seperate HE-muons from background events and unfold their energy spectrum.
The reconstructed spectrum of HE-muons is compatible with theoretical predictions and with previous results from the IceCube collaboration.
This spectrum does not include systematic effects compared to the published results shown in the gray band.
Therefore it is expected that the uncertainty of the reconstructed spectrum will increase with the inclusion of systematic effects.
The statistical significance is not sufficient to determine whether a prompt contribution is present on the predicted level in the resulting spectrum.

Since the uncertainty is due to the lack of simulated events, improvements of the final result are expected with growing numbers of simulations.
The available simulation statistics for the years $2012$ and above is $7$ times larger than in this analysis.
This will decrease the statistical uncertainty of the simulation and extend the results of this analysis to higher energies.

\bigskip 

\end{document}